\documentclass[%
 reprint,
superscriptaddress,
 noeprint,  
 nofootinbib, 
 amsmath,amssymb,
 aps,
]{revtex4-2}

\usepackage{graphicx}
\usepackage{dcolumn}
\usepackage{bm}

\usepackage{hyperref}
\usepackage{xcolor}
\hypersetup{%
    colorlinks={true},
    urlcolor={black}, 
    linkcolor={black},
    citecolor={black},
}
\usepackage{braket}

\usepackage[per-mode=symbol,separate-uncertainty]{siunitx}
\DeclareSIUnit{\dBm}{dBm}
\DeclareSIUnit{\dB}{dB}
\DeclareSIUnit{\sample}{Sa}

\newcommand{\rb}{\mathrm{b}}
\newcommand{\rd}{\mathrm{d}}
\newcommand{\re}{\mathrm{e}}
\newcommand{\rg}{\mathrm{g}}
\newcommand{\rp}{\mathrm{p}}
\newcommand{\qq}{\mathrm{q}} 
\newcommand{\rr}{\mathrm{r}}

\newcommand{\rx}{\mathrm{x}}
\newcommand{\rN}{\mathrm{N}}
\newcommand{\RO}{\mathrm{RO}}

\begin{document}
\title{Single-Shot Readout of a Superconducting Qubit Using a Thermal Detector}
\newcommand{\affilqcd}{QCD Labs, QTF Centre of Excellence, Department of Applied Physics, Aalto University, P.O. Box 13500, FIN-00076 Aalto, Finland}
\newcommand{\affiliqm}{IQM, Keilaranta 19, FI-02150 Espoo, Finland}
\newcommand{\affilvtt}{QTF Centre of Excellence, VTT Technical Research Centre of Finland Ltd., P.O.~Box 1000, 02044 VTT, Finland}
\author{András M. Gunyhó} \email{andras.gunyho@aalto.fi}\affiliation{\affilqcd}
\author{Suman Kundu} \affiliation{\affilqcd}
\author{Jian Ma} \affiliation{\affilqcd}
\author{Wei Liu} \affiliation{\affilqcd} \affiliation{\affiliqm}
\author{Sakari Niemelä} \affiliation{\affilqcd}
\author{Giacomo Catto} \affiliation{\affilqcd}
\author{Vasilii Vadimov} \affiliation{\affilqcd}
\author{Visa Vesterinen} \affiliation{\affilvtt}
\author{Priyank Singh} \affiliation{\affilqcd}
\author{Qiming Chen} \affiliation{\affilqcd}
\author{Mikko Möttönen} \affiliation{\affilqcd}\affiliation{\affilvtt}

\date{\today}

\begin{abstract}
    Measuring the state of qubits is one of the fundamental operations of a quantum computer.
    Currently, state-of-the-art high-fidelity single-shot readout of superconducting qubits relies on parametric amplifiers at the millikelvin stage.
    However, parametric amplifiers are challenging to scale beyond hundreds of qubits owing to practical size and power limitations.
    Nanobolometers have a multitude of properties that are advantageous for scalability and have recently shown sensitivity and speed promising for qubit readout, but such thermal detectors have not been demonstrated for this purpose.
    In this work, we utilize an ultrasensitive bolometer in place of a parametric amplifier to experimentally demonstrate single-shot qubit readout.
    With a readout duration of 13.9~\textmu{}s, we achieve a single-shot fidelity of 0.618 which is mainly limited by the energy relaxation time of the qubit, $T_1 = 28$~\textmu{}s.
    Without the $T_1$ errors, we find the fidelity to be 0.927.
    In the future, high-fidelity single-shot readout may be achieved by straightforward improvements to the chip design and experimental setup, and perhaps most interestingly by the change of the bolometer absorber material to reduce the readout time to the hundred-nanosecond level and beyond.
\end{abstract}

\maketitle

Qubit readout is a fundamental operation in quantum computing, both for determining the result at the end of a computation~\cite{divincenzo2000,nielsen2011}, as well as for error correction, which is necessary for fault-tolerance~\cite{shor1995,preskill1997,terhal2015}.
Currently, one of the most promising platforms for realizing a practically useful quantum computer is based on superconducting qubits~\cite{blais2004,barends2014,kjaergaard2020}.
Importantly, readout is currently one of the main bottlenecks on the way to quantum-error-corrected large-scale superconducting quantum processors: most of the error in the error correction cycles demonstrated in recent state-of-the-art experiments~\cite{zhao2022,acharya2023} arose during the readout phase. Thus improving existing readout techniques and especially discovering completely new advantageous ways to carry out readout are of great and urgent importance.

The standard method of measuring the state of superconducting qubits in the framework of circuit quantum electrodynamics is referred to as \emph{dispersive readout}~\cite{blais2004,koch2007,blais2021}.
Here, a qubit is dispersively coupled to a readout resonator, the frequency of which shifts depending on the qubit state.
The current state of the art in dispersive readout is a single-shot readout fidelity in excess of 99\% with an averaging time of less than \SI{100}{\nano\second} for single qubits~\cite{walter2017,sunada2022}, and 97--98\% on average for the simultaneous multiplexed readout of several qubits~\cite{heinsoo2018,acharya2023}.

To achieve a signal-to-noise ratio (SNR) sufficient for high-fidelity single-shot dispersive readout, the output signal from the readout resonator is typically amplified at the millikelvin stage by a parametric amplifier, such as a Josephson parametric amplifier (JPA)~\cite{aumentado2020} or a traveling-wave parametric amplifier (TWPA)~\cite{esposito2021}.
These amplifiers can be quantum limited, i.e., the noise added by the amplification stems solely from the Heisenberg uncertainty principle for the in-phase and quadrature components of the amplified signal~\cite{clerk2010,bergeal2010,macklin2015}.
Parametric amplifiers are widely utilized due to their high gain and low noise, but they suffer from some drawbacks that challenge their applicability when scaling to large numbers of qubits.
Namely, JPAs have a fairly narrow bandwidth, which renders them less suitable for multiplexed qubit readout~\cite{heinsoo2018}.
On the other hand, although TWPAs provide a broad bandwidth, they have limited dynamic range for the purposes of highly multiplexed readout and typically incorporate more than $10^3$ Josephson junctions, rendering their on-chip footprint sizeable and high-yield large-scale fabrication challenging.
Importantly, both of these amplifier types require strong isolation between the amplifier and the qubit-resonator system.
In the case of JPAs, this is because the amplifier works by reflection, requiring the use of a circulator, and TWPAs require a relatively strong pumping tone possibly near the frequencies of the qubit and the readout resonator.
Furthermore, TWPAs amplify vacuum and other noise, which may be reflected by later components in the amplification chain due to imperfect impedance matching. This noise may leak backwards to the TWPA input and cause decoherence in the qubit if not isolated.
To mitigate these issues, permanent-magnet microwave isolators are placed between the readout transmission line and the amplifier input.
Such isolators introduce losses in the signal, are very large in size, costly, and require bulky shielding to protect both the qubit and the amplifier from the magnetic fields they introduce.

These scalability issues have motivated the development of novel readout techniques that evade the need for parametric amplifiers.
For example, Ref.~\cite{opremcak2021} introduces a microwave photon counter instead of voltage amplification at the millikelvin stage.
However, this method requires a more involved pulse sequence, and it suffers from the backaction and possible creation of quasiparticles due to their tunneling events.
In Ref.~\cite{chen2022}, qubit readout is achieved without a parametric amplifier by driving the qubit to the second excited state before readout, at the cost of an increased measurement time. However, the used non-parametric microwave amplifier introduces high-temperature noise close to the qubit frequency, which calls for bulky microwave isolators.
Another currently developing approach to scalability is to deliver signals to and from the cryostat in the optical domain~\cite{lecocq2021,delaney2022}.

Interestingly, ultrasensitive nanobolometers~\cite{govenius2016,kokkoniemi2019} have been recently demonstrated to be fast and sensitive enough for readout of superconducting qubits, reaching thermal time constants in the hundred nanosecond scale and energy resolution of a few typical microwave photons~\cite{kokkoniemi2020}.
Such bolometers have a number of characteristic properties that are attractive for scalable qubit readout.
In contrast to a parametric amplifier, the bolometer introduced in Ref.~\cite{govenius2016} is driven by a probe tone with a frequency below \SI{1}{\giga\hertz}, well below the typical frequencies 5--\SI{8}{\giga\hertz} of the qubit and the resonator.
In addition, the bolometer can be probed with a low power of roughly \SI{-130}{\dBm} at the chip, while for example TWPAs typically require pump powers of \SI{-75}{\dBm} or higher~\cite{esposito2021}.
Furthermore, the absorbing port of the bolometer can be conveniently matched to $\SI{50}{\ohm}$~\cite{kokkoniemi2019}.
This presents a cold bath to the readout transmission line, while allowing for the detection frequency to vary by orders of magnitude given a fixed probe frequency.
These features may eliminate the need for isolators between the qubit-resonator system and the bolometer, and together with the small size of the bolometers render them highly promising from the scalability point of view.

Another advantage of bolometers compared with parametric amplifiers is that since a bolometer measures power, or photon number, it is not bound to add quantum noise stemming from the Heisenberg uncertainty principle.
The vacuum noise does not promote detection events in the bolometer since no energy can be extracted from the vacuum. Thus bolometric readout is fundamentally different from the usual dispersive readout.

In addition, bolometers are relatively simple to both fabricate and operate~\cite{kokkoniemi2019}.
They do not require engineering of a large number of small Josephson junctions, and a bolometer requires only a single continuous probe tone with two parameters, power and frequency, to optimize performance.

Owing to these appealing features, it is of great interest to study whether the fundamentally different operation principle of bolometers can be harnessed in the readout of superconducting qubits.
However, no thermal detector to date has been employed for this purpose.

In this work, we integrate an ultrasensitive bolometer at millikelvin temperature to the readout circuitry of a superconducting qubit (Fig.~\ref{fig:experiment_schematic}). After the characterization of the bolometer and the qubit (Fig.~\ref{fig:example_measurements}), we demonstrate single-shot qubit readout at the fidelity of 0.618 (Fig.~\ref{fig:singleshot_histogram}). Taken that we have not used the fastest compatible bolometers~\cite{kokkoniemi2020}, but those that are orders of magnitude slower~\cite{kokkoniemi2019}, this first demonstration of bolometric single-shot qubit readout seems a promising milestone for future high-fidelity scalable readout of superconducting qubits.

\subsection*{Experimental setup}

\begin{figure}
    \includegraphics[width=\linewidth]{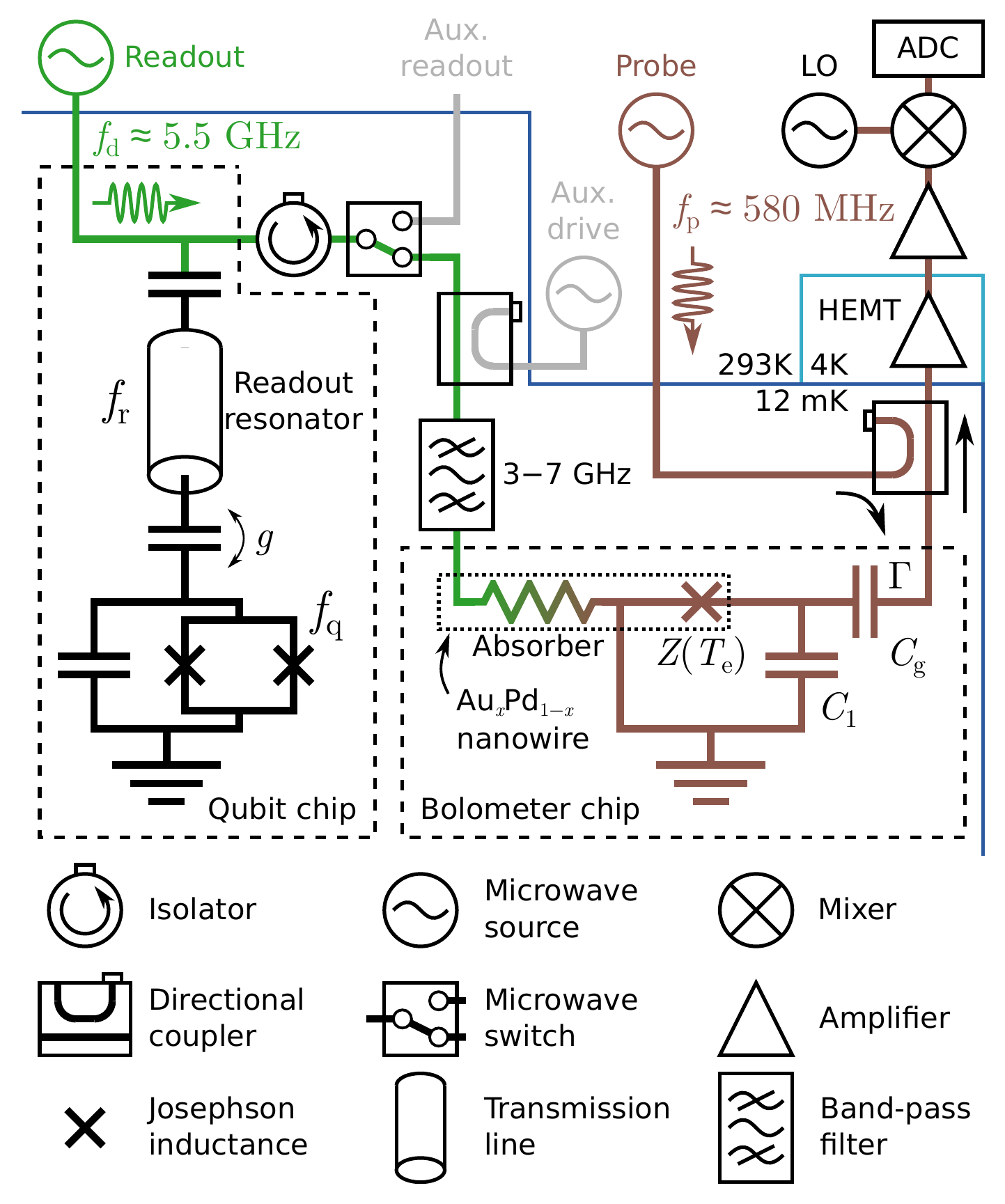}
    \caption{%
        \label{fig:experiment_schematic}%
        \textbf{Experimental setup.}
        The readout resonator is driven at frequency $f_\rd\approx\SI{5.5}{\giga\hertz}$ by a readout tone which is reflected off of the resonator and absorbed by a metallic nanowire in the absorber port of the bolometer.
        The absorbed radiation shifts the electron temperature $T_{\re}$ and consequently the impedance $Z(T_{\re})$ of a chain of SNS junctions, marked by a brown cross.
        A continuous probe tone with frequency $f_\rp\approx\SI{580}{\mega\hertz}$ is reflected off of the gate capacitor $C_{\rg}$ of the bolometer as determined by the reflection coefficient $\Gamma$.
        The reflected signal is amplified, digitized, and used to determine the qubit state.
    }
\end{figure}

A schematic diagram of the experimental setup is shown in Fig.~\ref{fig:experiment_schematic}.
A standard flux-tunable Xmon qubit~\cite{barends2013} with a transition frequency $f_\qq =\SI{7.655}{\giga\hertz}$ at the flux sweet spot, anharmonicity $\alpha/(2\pi) = \SI{-273}{\mega\hertz}$,
energy relaxation time $T_1 = \SI{28}{\micro\second}$, and Ramsey dephasing time $T_2 = \SI{7.6}{\micro\second}$ is capacitively coupled to a coplanar readout resonator in a notch configuration.
The resonator has its fundamental resonance frequency at $f_{\rr,\rg} = \SI{5.473}{\giga\hertz}$ if the qubit is in its ground state $\ket{\rg}$ and a linewidth $\kappa_{\rr}/(2\pi) = \SI{1.0}{\mega\hertz}$.
The resonator is capacitively coupled to the qubit with a coupling strength $g/(2\pi) = \SI{61}{\mega\hertz}$.
These parameters produce a dispersive shift of $\chi/2\pi = -\SI{0.3}{\mega\hertz}$, i.e., the resonator frequency shifts to $f_{\rr, \re} = f_{\rr, \rg} + 2\chi/(2\pi)$ if the qubit is in its excited state $\ket{\re}$.

To read out the qubit state, a rectangular microwave pulse of length $t_{\RO} \approx \SI{10}{\micro\second}$, frequency $f_\rd$, and power $P_{\rd}$ is applied to the feedline of the readout resonator.
In typical dispersive readout, the drive frequency is chosen to be in the middle of the dressed resonator frequencies, $f_{\rd} = (f_{\rr,\rg} + f_{\rr,\re})/2$, and the photons reflected from the resonator accumulate a phase shift depending on the qubit state~\cite{blais2004}.
In contrast, here we operate in a photodetection mode~\cite{nesterov2020} by driving close to one of the dressed frequencies, $f_{\rd} \approx f_{\rr,\rg}$.
In this driving scheme, information about the qubit state is mostly carried by the power of the signal, i.e., the number of photons emitted by the resonator into the feedline.
A power difference is necessary for the bolometer, since it is a thermal detector, sensitive to the power but insensitive to the phase of its input signal.
Note that for the standard dispersive readout, the ratio $|\chi|/\kappa_\rr = 1/2$ yields the optimal SNR~\cite{gambetta2008}, while for photodetection-based readout, the optimal ratio is higher~\cite{nesterov2020}.

The output of the feedline for the readout resonator is connected to the absorber port of the bolometer.
The bolometer, similar to the device of Ref.~\cite{kokkoniemi2019}, resides on a chip different from that of the qubit-resonator system as shown in Fig.~\ref{fig:experiment_schematic}.
The main component of the bolometer is a resistive $\textnormal{Au}_{x}\textnormal{Pd}_{1-x}$ ($x \approx 0.6$) nanowire.
A segment of the nanowire works as the resistive absorber and the remaining wire is interrupted by a series of superconducting Al islands, forming a chain of SNS junctions.
The impedance $Z(T_{\re})$ of this junction chain depends on the electron temperature of the nanowire $T_{\re}$.
The junction chain is embedded in an effective $LC$ circuit formed by a shunt capacitor $C_1 = \SI{134}{\pico\farad}$ in parallel with $Z(T_{\re})$, which can be modeled as a parallel resistance and inductance~\cite{govenius2016}.
The nanowire is grounded between the absorber and the junctions, so that the essentially purely real-valued impedance of the absorber does not contribute to the $LC$ circuit.

The bolometer is probed by reflecting a continuous tone at power $P_{\rp}$ and frequency $f_{\rp}$ from the gate capacitor $C_{\rg} = \SI{0.87}{\pico\farad}$.
With a low $P_{\rp}$, the $LC$ circuit resonates at the frequency $f_{\rb} = \SI{585}{\mega\hertz}$ with a linewidth $\SI{7.6}{\mega\hertz}$.
As radiation is absorbed by the absorber, $T_{\re}$ increases, which shifts $f_{\rb}$ down, and thus by observing changes in the reflection coefficient $\Gamma$ at the gate capacitor, it is possible to detect the radiation incident on the bolometer input.
The reflected probe signal is amplified by a low-noise high-electron-mobility-transistor (HEMT) amplifier at $\SI{4}{\kelvin}$, and further amplified, demodulated and digitized at room temperature in a heterodyne configuration.

\begin{figure}
    \includegraphics[width=\linewidth]{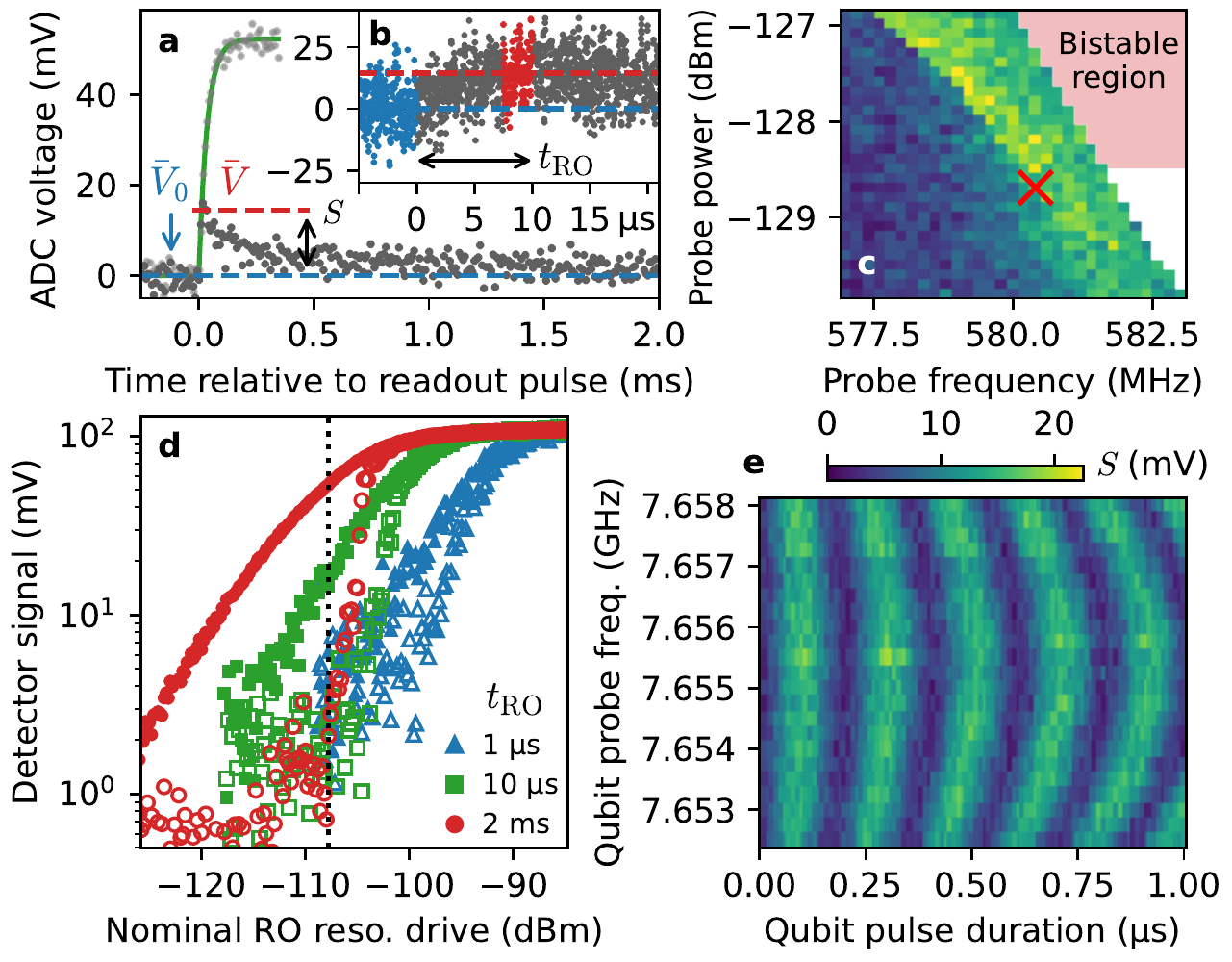}
    \caption{%
        \label{fig:example_measurements}%
        \textbf{Characterization experiments.}
        (a) Example timetrace of the probe signal reflected from the bolometer with a qubit readout pulse length of $t_{\RO} = \SI{10}{\micro\second}$ (dark gray dots) and $>\SI{1}{\milli\second}$ (light gray dots).
        The solid green line is an exponentially rising fit to the long pulse, and the dashed horizontal lines indicate the extracted values of $\bar{V}$ and $\bar{V}_0$ used to define the detector signal $S$ for the short pulse.
        The parameter values in all panels are as shown in Table~\ref{table:characterization_parameters} unless otherwise specified.
        (b) As (a) but only for the \SI{10}{\micro\second} readout pulse.
        The blue and red-colored regions indicate where $\bar{V}_{0}$ and $\bar{V}$ are averaged, respectively.
        These data are taken with 128 ensemble averages.
        The signal appears noisier than in panel (a), since each data point in (a) is calculated by averaging 512 adjacent data points.
        (c) Detector signal $S$ as a function of the probe frequency $f_{\rp}$ and probe power $P_{\rp}$.
        In the region shaded with red, where $P_{\rp} \gtrsim \SI{-128.5}{\dBm}$, the bolometer exhibits bistability due to electrothermal feedback.
        The red cross indicates the chosen operation point for qubit readout.
        The colorbar is shared with panel (e).
        (d)~Detector signal as a function of the nominal power of the readout pulse $P_{\rd}$ for various indicated readout pulse lengths and for the readout pulse applied off-resonance ($f_{\rd} = \SI{5.400}{\giga\hertz}$, filled markers) and on-resonance ($f_{\rd} = \SI{5.473}{\giga\hertz} = f_{\rr,\rg}$, unfilled markers).
        The resonator becomes significantly nonlinear at approximately \SI{-108}{\dBm}, indicated by the black dotted vertical line.
        (e) Detector signal as a function of the resonant qubit drive pulse length and frequency, showing Rabi oscillation.
        These data are taken with 512 ensemble averages.
    }
\end{figure}

The shift in $f_{\rb}$ due to the readout pulse incoming from the readout resonator is observed as a change in the digitized voltage [see Fig.~\ref{fig:example_measurements}(a)--(b)].
For low $P_{\rp}$, the readout pulse causes $T_{\re}$ and thus the reflected signal to approach a steady state value exponentially with a thermal time constant $\tau_{\rb}$.
The time constant depends on $f_{\rp}$, $P_{\rp}$, and the power of the readout pulse $P_{\rd}$ in a non-trivial way due to electrothermal feedback~\cite{govenius2016,kokkoniemi2019}.
For the relevant parameter regime considered here, $\tau_{\rb}$ varies between $\SI{10}{\micro\second}$ and $\SI{1}{\milli\second}$.
Notably, we have $\tau_{\rb} \gtrsim T_1$, which implies that for qubit readout, we must operate the bolometer in a calorimetric fashion, i.e., with $t_{\RO} < \tau_{\rb}$.
This is highlighted in Fig.~\ref{fig:example_measurements}(a), where the time constant, extracted from a measurement with a long readout pulse $t_{\RO} > \SI{1}{\milli\second}$, is $\tau_{\rb} = \SI{36.2}{\micro\second}$.
With $t_{\RO} = \SI{10}{\micro\second}$, which is more feasible for qubit readout than $t_{\RO} > \tau_{\rb}$, the steady state is far from the reached maximum signal level.

For high enough $P_{\rp}$, the electrothermal feedback results in a bistability for the electron temperature of the SNS junction.
This bistability can be exploited for high-fidelity photodetection by operating the bolometer in a latching mode using a pulsed probe tone~\cite{govenius2016}.
However, this scheme introduces significant dead time for the detection, and requires detailed calibration of the pulse shape.
For simplicity, we thus focus on the continuous detection scheme with sufficiently low $P_{\rp}$.

Using time-domain data digitized from the bolometer output, we define the detector signal as
\begin{align}
    \label{eq:signal}
    S = \bar{V} - \bar{V}_0,
\end{align}
where $\bar{V}_0$ is the time-average of the digitized voltage before the readout pulse and $\bar{V}$ is the time-average of the voltage over some averaging window $[t_0, t_\RO]$.
Since $t_{\RO} < \tau_{\rb}$, increasing the time over which the average is taken can significantly decrease $\bar{V}$, but may increase the SNR.
In practice, we observe that choosing $t_0 = 0.75\times t_{\RO}$ balances these two effects reasonably well.

Both the qubit-resonator and bolometer chips are mounted on their individual sample holders, which are placed in separate magnetic shields and attached to the mixing-chamber plate of a dilution refrigerator with the base temperature of \SI{12}{\milli\kelvin}.
Note that in the setup presented in Fig.~\ref{fig:experiment_schematic}, an additional microwave switch and directional coupler are placed between the qubit and bolometer.
These allow measuring the qubit and driving the bolometer individually, and are only used for initial separate characterization of the qubit and the bolometer.
The auxiliary readout channel called for the use of an isolator between the chips, but this can be removed in future experiments.

\section*{Results}

During initial characterization, we apply no driving to the readout resonator of the qubit and find the bolometer resonance by measuring the reflection coefficient of the bolometer probe signal as a function of $f_{\rp}$ and $P_{\rp}$.
Next, we apply pulses to the readout resonator with $f_{\rd}$ well detuned from $f_{\rr,\rg}$, and map the detector signal $S$ as a function of $f_{\rp}$ and $P_{\rp}$.
These results, shown in Fig.~\ref{fig:example_measurements}(c), provide us with a feasible operation point $(f_{\rp}, P_{\rp}) = (\SI{580.5}{\mega\hertz}, \SI{-128.7}{\dBm})$ where $S$ is maximized with $P_{\rp}$ below the region where the electrothermal feedback induces bistability as discussed above.
Table~\ref{table:characterization_parameters} summarizes the parameter values used during the initial characterization.

\begin{table}
    \caption{Typical parameter values used during characterization measurements. Note that all powers we report in this manuscript are uncalibrated, nominal values reaching the corresponding chip, based on the estimated attenuation of the lines in our setup.}
    \begin{tabular}{c|c|c}
        Quantity & Symbol & Value \\ \hline
        Bolometer probe frequency & $f_{\rp}$ & \SI{580.5}{\mega\hertz} \\
        Bolometer probe power (nominal) & $P_{\rp}$ & \SI{-128.7}{\dBm} \\
        Readout resonator drive frequency & $f_{\rd}$ & \SI{5.400}{\giga\hertz} or $f_{\rr,\rg}$ \\
        Readout drive power (nominal) & $P_{\rd}$ & \SI{-107.8}{\dBm} \\
        Readout pulse length & $t_{\RO}$ & \SI{10}{\micro\second} \\
        Number of ensemble-averages & & 16
    \end{tabular}
    \label{table:characterization_parameters}
\end{table}

With the bolometer operation point fixed, we carry out standard qubit characterization measurements~\cite{chen2018} while monitoring the bolometer signal.
For example, Fig.~\ref{fig:example_measurements}(d) shows $S$ as a function of the qubit readout power $P_{\rd}$, both on resonance $f_{\rd} = f_{\rr,\rg}$ and off-resonance at $f_{\rd} = f_{\rr, \rg} - \SI{70}{\mega\hertz}$, with the qubit drive turned off.
Using $P_{\rd} = \SI{-126}{\dBm}$, a maximal contrast of $\SI{32}{\dB}$ between the on and off-resonance drives is achieved with a very long readout pulse.
More importantly, Fig.~\ref{fig:example_measurements}(d) further highlights that the long bolometer time constant is limiting the readout signal since the contrast is greatly reduced for the microsecond pulses, which are still roughly an order of magnitude longer than the current state-of-the-art qubit readout.
A reasonable contrast is not achieved with a $\SI{1}{\micro\second}$ readout pulse even for the highest feasible readout power of $\SI{-108}{\dBm}$, beyond which the resonator becomes nonlinear and the quantum non-demolition nature of the readout breaks down~\cite{cohen2022,khezri2022}.

After finding the resonator and qubit frequencies using the standard single-tone and two-tone spectroscopies adjusted to our bolometric readout, we carry out a Rabi oscillation measurement.
We initialize the qubit to its ground state, drive it with a rectangular pulse, and measure the readout resonator using the bolometer. Here, we employ a $\SI{10}{\micro\second}$ readout pulse and ensemble-average the result 512 times.
With the length and the modulation frequency of the qubit drive pulse are varied, we obtain a clear chevron pattern as desired [see Fig.~\ref{fig:example_measurements}(e)], from which we determine a $\pi$ pulse length of $\SI{100}{\nano\second}$.

\begin{figure}
    \includegraphics[width=\linewidth]{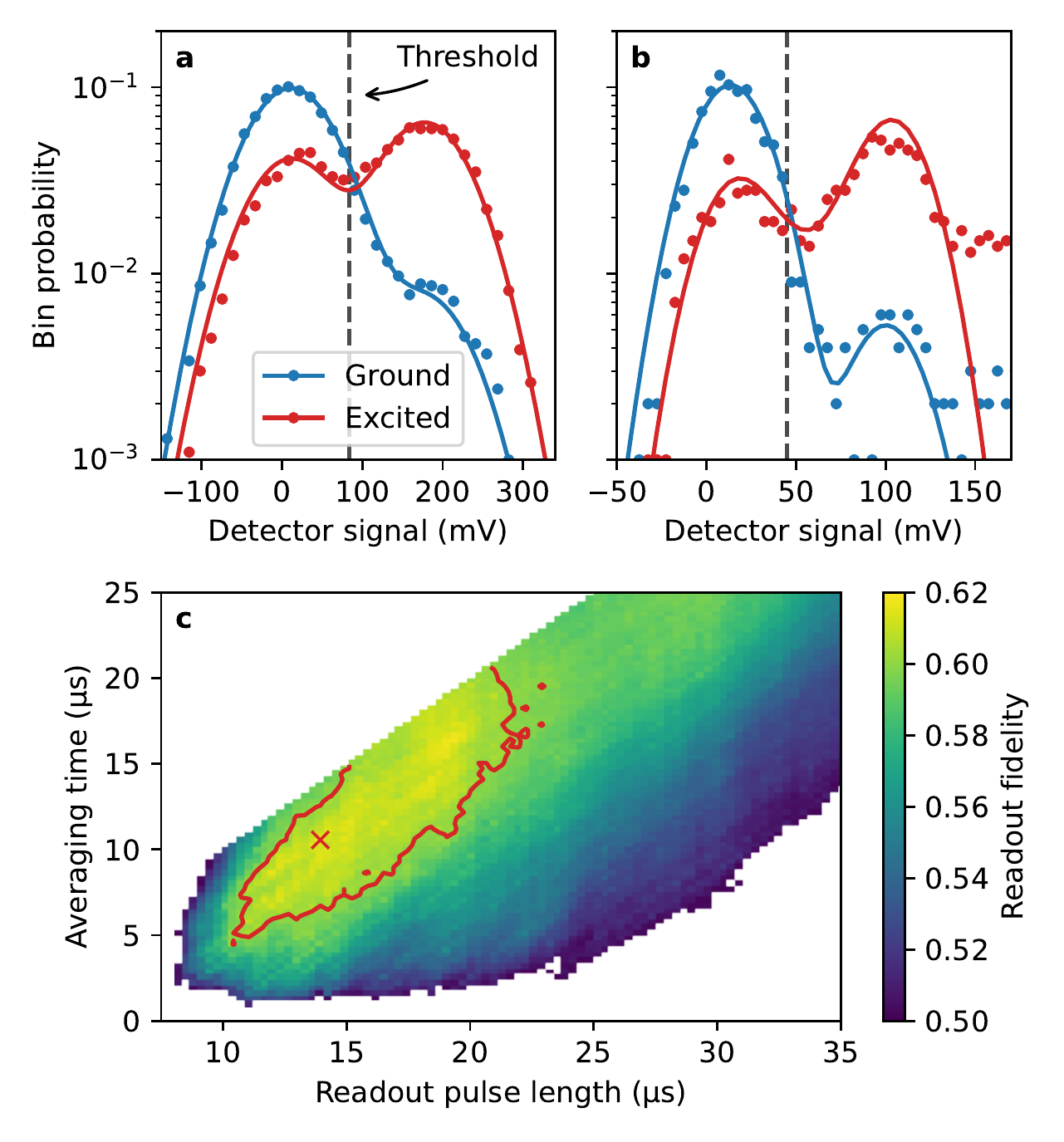}
    \caption{%
        \label{fig:singleshot_histogram}%
        \textbf{Single-shot readout.}
        (a) Probability distributions of the single-shot qubit readout signal $S$ for the qubit prepared in the ground (blue dots) and excited (red dots) states.
        The solid lines are fits to a sum of two Gaussian distributions.
        The vertical dashed line indicates the threshold which yields the highest readout fidelity.
        The bolometer is tuned to its operation point and the readout power is $P_{\rd} = \SI{-108}{\dBm}$ with a pulse length of $t_{\RO} = \SI{20}{\micro\second}$ and an averaging time of \SI{5}{\micro\second} for $\bar V$.
        (b) As (a) but for $t_{\RO} = \SI{13.9}{\micro\second}$ and an averaging time of \SI{10.6}{\micro\second} for $\bar V$ and long averaging for $\bar V_0$ (see text).
        Here, the fit for the excited state is obtained using a model that takes into account the $T_1$ decay of the qubit during the readout (see Methods).
        (c) Single-shot readout fidelity as a function of the readout pulse length and averaging time for $\bar V$, both changed in post processing.
        The red boundary indicates the region where the fidelity is greater than 0.6.
        The highest fidelity is achieved with the parameters indicated by the red cross, where the data of panel (b) are obtained.
        }
\end{figure}

With the qubit frequency and $\pi$ pulse length calibrated, we carry out single-shot qubit readout by alternating between preparing the qubit in $\ket{\rg}$ and $\ket{\re}$, and recording $S$ with no ensemble-averaging.
For each prepared state, we record $10^4$ data points.
The data are binned to produce a histogram, with the bin width chosen using Scott's rule~\cite{scott1979}.
To maximize the readout fidelity $F = 1 - P(\rg | \re) - P(\re | \rg)$, we optimize the threshold value for $S$ to decide the measurement outcome, $\ket{\rg}$ or $\ket{\re}$.
Here, $P(a|b)$ is the probability of measuring the qubit in the state $\ket{\textrm{a}}$ provided that it was prepared in the state $\ket{\textrm{b}}$, where the probabilities are obtained from the measured distributions.
The highest fidelity is obtained by increasing the readout pulse length to $t_{\RO} = \SI{20}{\micro\second}$, so that the above-discussed choice of $t_0 = 0.75\times t_{\RO}$ yields an averaging time of $\SI{5}{\micro\second}$ for $\bar{V}$, and setting the readout power to $P_{\rd} = \SI{-108}{\dBm}$, which is just below the point of nonlinearity for the readout resonator.
Figure~\ref{fig:singleshot_histogram}(a) shows the measured probability distributions for the probe signal $S$ with these parameters, from which we extract the fidelity $F = 0.49$.

In an effort to further optimize the single-shot readout fidelity, we carry out an additional experiment, where we intentionally stretch the readout pulse to be unreasonably long, $\SI{40}{\micro\second}$, and instead of just storing $S$ derived from the time-averaged quantities $\bar{V}_0$ and $\bar{V}$, we record the full time traces of the bolometer output signal for 1000 single shots.
With these data, we may vary the effective readout pulse length and averaging time of $\bar{V}$ in post processing.
Figure~\ref{fig:singleshot_histogram}(c) shows the resulting single-shot readout fidelity as a function of the digitally determined pulse length and averaging time.
We observe a region of relatively high fidelity $F > 0.6$ [highlighted by the red boundary in Fig.~\ref{fig:singleshot_histogram}(c)], with the highest fidelity of 0.618 achieved with $t_{\RO} = \SI{13.9}{\micro\second}$ and $t_{\RO} - t_0 = \SI{10.6}{\micro\second}$.
Figure~\ref{fig:singleshot_histogram}(b) shows the probability distributions of the signal with these parameter values.

The fidelity is overall higher in Fig.~\ref{fig:singleshot_histogram}(b)--(c) than in Fig.~\ref{fig:singleshot_histogram}(a).
This is mostly because in panel (a), $\bar V_0$ is calculated individually for each single shot.
Since each value of $\bar V_0$ is averaged over a relatively short, \SI{1.1}{\micro\second} window, it introduces significant noise to the value of $S$ calculated using Eq.~\eqref{eq:signal}.
In contrast, a common value of $\bar V_0$ is used for the whole data set of panels (b)--(c), meaning that only the fluctuations of $\bar V$ contribute to the noise of $S$.
Rescaling the standard deviation of the data of panels (b)--(c) by a factor corresponding to the noise from this \SI{1.1}{\micro\second} window reproduces the increased noise level of panel (a).

For the qubit prepared in $\ket{\re}$, the single-shot probability distribution is clearly bimodal in Figs.~\ref{fig:singleshot_histogram}(a)--(b).
This is expected, since $t_{\RO}$ is comparable to the qubit energy decay time $T_1$, and thus the qubit experiences significant spontaneous relaxation during the readout.
Even in the case of infinitely narrow distributions and ideal state preparation, the $T_1$ decay produces an error of approximately $1 - \exp\left[-t_{\RO} / (2T_1)\right] \approx 22\%$ for $t_{\RO} = \SI{13.9}{\micro\second}$ and $T_1 = \SI{28}{\micro\second}$~\cite{gambetta2007}.
The long readout is stems from the relatively long effective thermal time constant of the bolometer, $\tau_{\rb} = \SI{9.4}{\micro\second}$, extracted from the data of Fig.~\ref{fig:singleshot_histogram}(c) for the qubit prepared in $\ket{\rg}$. In Fig.~\ref{fig:singleshot_histogram}(c),
the trend of relatively high and constant fidelity obtained for simultaneously increasing readout pulse length and averaging time indicates that the increase in the fidelity owing to the increasing bolometer SNR  approximately compensates the decrease in the fidelity owing to increasing qubit decay.
With short readout times, the fidelity is low because the distributions corresponding to the different qubit states are not well separated.

The data of Fig.~\ref{fig:singleshot_histogram}(a), as well as the $\ket{\rg}$ state data of Fig.~\ref{fig:singleshot_histogram}(b) are well modeled by a sum of two Gaussian distributions.
For the $\ket{\re}$ state data of Fig.~\ref{fig:singleshot_histogram}(b), we utilize a  model that incorporates the decay of the qubit during the readout (see Methods).
This advanced model is not needed for the $\ket{\re}$ state data of Fig.~\ref{fig:singleshot_histogram}(a), since the effect of the decay on the distribution is masked by noise.
Using the parameters extracted from these fits, we determine the fidelity with $t_{\RO} = \SI{13.9}{\micro\second}$ and the $T_1$ error removed to be $F \approx 0.927$.
The $\ket{\re}$ state data in Fig.~\ref{fig:singleshot_histogram}(b) deviate from the fit above roughly \SI{150}{\milli\volt}.
We attribute this to qubit excitations outside the computational subspace owing to high readout power.

\section*{Conclusions}

In summary, we demonstrated that a state-of-the-art thermal detector is capable of achieving reasonable-fidelity single-shot readout of a superconducting qubit, i.e., no voltage amplification is required for the qubit readout signal.
We were able to carry out qubit characterization measurements with significantly less ensemble-averaging than in previous setups lacking millikelvin amplifiers.
Using a \SI{13.9}{\micro\second} readout pulse, we achieve a single-shot readout fidelity of 0.62, and a fidelity of 0.93 after removing errors arising from finite qubit lifetime.

Several factors in our experiment can be improved to increase the single-shot readout fidelity.
Most significantly, the SNR in the single-shot readout is fairly low, which mainly arises from the long effective thermal time constant of the bolometer, which is on the order of ten microseconds.
This motivates to use a long readout pulse, which degrades the fidelity due to the $T_1$ decay of the qubit.
A straightforward way to reduce the time constant is to use a material with a lower heat capacity in place of AuPd as the absorber.
A promising candidate for this is graphene, which has been observed to lead to extremely high-sensitivity bolometers with time constants in the hundred-nanosecond range~\cite{mckitterick2013,kokkoniemi2020}.

Apart from the bolometer chip itself, minor modifications to our experimental setup may significantly improve the SNR.
In the experiments presented here, we have a number of additional microwave components between the qubit and bolometer, which were used to independently characterize the qubit and the bolometer.
Removing these components and directly connecting the output of the qubit readout feedline to the absorber of the bolometer removes the loss of signal between the two chips.
Ultimately, the bolometer could be directly connected to the readout feedline, either on-chip or in a layered flip-chip architecture~\cite{kosen2022}.
The losses between the qubit and bolometer are an important factor in degrading the fidelity especially in our case, since the maximum power we can apply to the resonator is on the low end of what the bolometer can detect.

It is also possible to optimize the qubit chip design to be better suited for photodetection-based readout.
In our particular design, the readout resonator is placed in a notch-type configuration with symmetric coupling to the input and output feedlines.
This is disadvantageous, since effectively half of the signal reflected from the resonator is lost as it escapes through the input port.
Changing to a transmission configuration with a weak coupling to the input port is expected to significantly increase the power incident on the bolometer.
The ratio $|\chi| / \kappa_\rr = 0.3$ in our sample is close to the value of $1/2$ which maximizes the SNR in typical dispersive readout, but it is sub-optimal for photodetection-based readout where a larger ratio is desirable. Furthermore, our readout resonator lies below the qubit frequency which leads to the onset of non-adiabatic and even chaotic resonator dynamics at a much lower photon number than for an elevated readout frequency~\cite{blais2021,cohen2022,khezri2022}. Thus designing the readout circuit to resonate above the qubit frequency can lead to a greatly increased signal power at the bolometer input.

Finally, the readout fidelity can be improved by advanced pulsing schemes and improved data analysis methods.
Instead of a simple rectangular pulse, the resonator can be driven by a two-step pulse~\cite{walter2017}, where the drive amplitude is initially high for a short amount of time.
This may enable the use of a power level that the bolometer can detect with higher SNR, still maintaining the quantum-non-demolition nature of the readout.
Alternatively, the SNR of the signal reaching the bolometer may be increased by preparing the qubit in a higher excited state prior to readout~\cite{chen2022}, or by using a two-tone drive that induces an effective longitudinal coupling between the qubit and the bolometer~\cite{touzard2019,ikonen2019}.
Data analysis can be improved by utilizing more sophisticated signal processing than the simple boxcar time-averaging employed here.
For example, the time-averaging can be weighted by the separation between the average trajectories corresponding to $\ket{\rg}$ and $\ket{\re}$~\cite{walter2017}, or the single-shot trajectories can be classified using a machine learning algorithm~\cite{lienhard2022}.

Note that in our current analysis, we have neglected the falling edge of the voltage signal that can be taken into use in the advanced analysis methods to increase the SNR. The averaging over the falling edge does not increase the $T_1$-related infidelity since during this averaging time, the readout pulse is off and hence possible qubit decay will not lead to a change in the bolometer signal. In fact, using a graphene bolometer in the calorimetric mode seems promising to improve the readout fidelity (Methods).

By incorporating the improvements discussed above, we estimate that it may be possible to carry out single-shot qubit readout at 99.9\% level of fidelity using bolometers (see Methods).
Thus, bolometers are a promising component for scalable high-fidelity readout of superconducting qubits owing to their low power consumption, large probe tone frequency offset from qubits, resilience against quantum noise, small footprint, no need for microwave isolators, and naturally introduced low-temperature bath for qubits that can be impedance matched at a very broad range of readout frequencies.

\renewcommand{\figurename}{\textbf{Extended Data Fig.}}
\setcounter{figure}{0}

\renewcommand{\tablename}{\textbf{Extended Data Table}}
\setcounter{table}{0}

\section*{Methods}

\subsection{Bolometer fabrication}

In the bolometer fabrication, we begin with a four-inch silicon wafer (100) of high resistivity ($\rho$ $>$ 10 k$\Omega$cm), covered by a 300-nm thermal oxide. Then, we sputter 200 nm of pure Nb onto the wafer. We use AZ5214E photoresist in positive mode with hard contact to define the waveguide in a Karl Suss MA-6 mask aligner. After development, the sample undergoes etching using a Plasmalab 80Plus Oxford Instruments RIE system. The plasma operates with a gas flow of $\mathrm{SF_{6}}$/$\mathrm{O_{2}}$ at 40 sccm/20 sccm with an rf power of 100~W.

We clean the resist residuals in acetone and IPA using an ultrasonic excitation, and then dry the chip with a nitrogen gun. Next, we use an atomic-layer deposition (ALD) method to grow a 45-nm dielectric layer of $\mathrm{Al_{2}O_{3}}$ in a Beneq TFS-500 system. Subsequently, we protect the dielectric layer at desired capacitor regions using AZ5214E resist, and wet-etch the rest of the ALD oxide with an ammonium fluoride–hydrofluoric acid mixture. We then cleave the 4-inch wafer into a 2 $\times$ 2 cm$^{2}$ chip using Disco DADdy.

The nanowire is patterned using EPBG5000pES electron beam lithography (EBL) using a bilayer of MMA/PMMA resist on a single chip. We deposit a 30-nm-thick AuPd layer in an electron beam evaporator at a rate of 0.5 Å/s. After liftoff in acetone overnight, we pattern the superconducting leads galvanically connected to the nanowire by EBL and deposit them with 100 nm of Al at a rate of 5 Å/s. Finally, we cleave each pixel (5 $\times$ 5 mm$^{2}$) using a laser micromachining system and package the chosen chip. We employ Al bonding wires to connect the chip to the printed circuit board of the sample holder.

\subsection{Qubit fabrication}

The qubit samples are fabricated using the following steps: First, a layer of 200 nm of Nb is sputtered onto a high-resistivity silicon substrate. Next, we define the transmission line, readout resonator, and transmon shunt capacitor using photolithography, followed by dry reactive ion etching (RIE).

The Al/Al$_2$O$_3$/Al Josephson junctions are subsequently fabricated using standard electron beam lithography and the Dolan bridge method, where oxidation is used between the deposition of two Al layers to form the Josephson junctions. To ensure galvanic contact between Al and Nb, the niobium oxide is removed by argon milling prior to the deposition of any Al.

The excess metal is lifted off in acetone. The room temperature resistance of the Al/Al$_2$O$_3$/Al junctions are measured to select a sample that is most likely to yield the desired qubit frequency. Finally, the chip is diced into individual samples and the selected sample is wire-bonded to a sample holder.

\subsection{Details of the experimental setup}

\begin{figure*}
    \includegraphics[width=\linewidth]{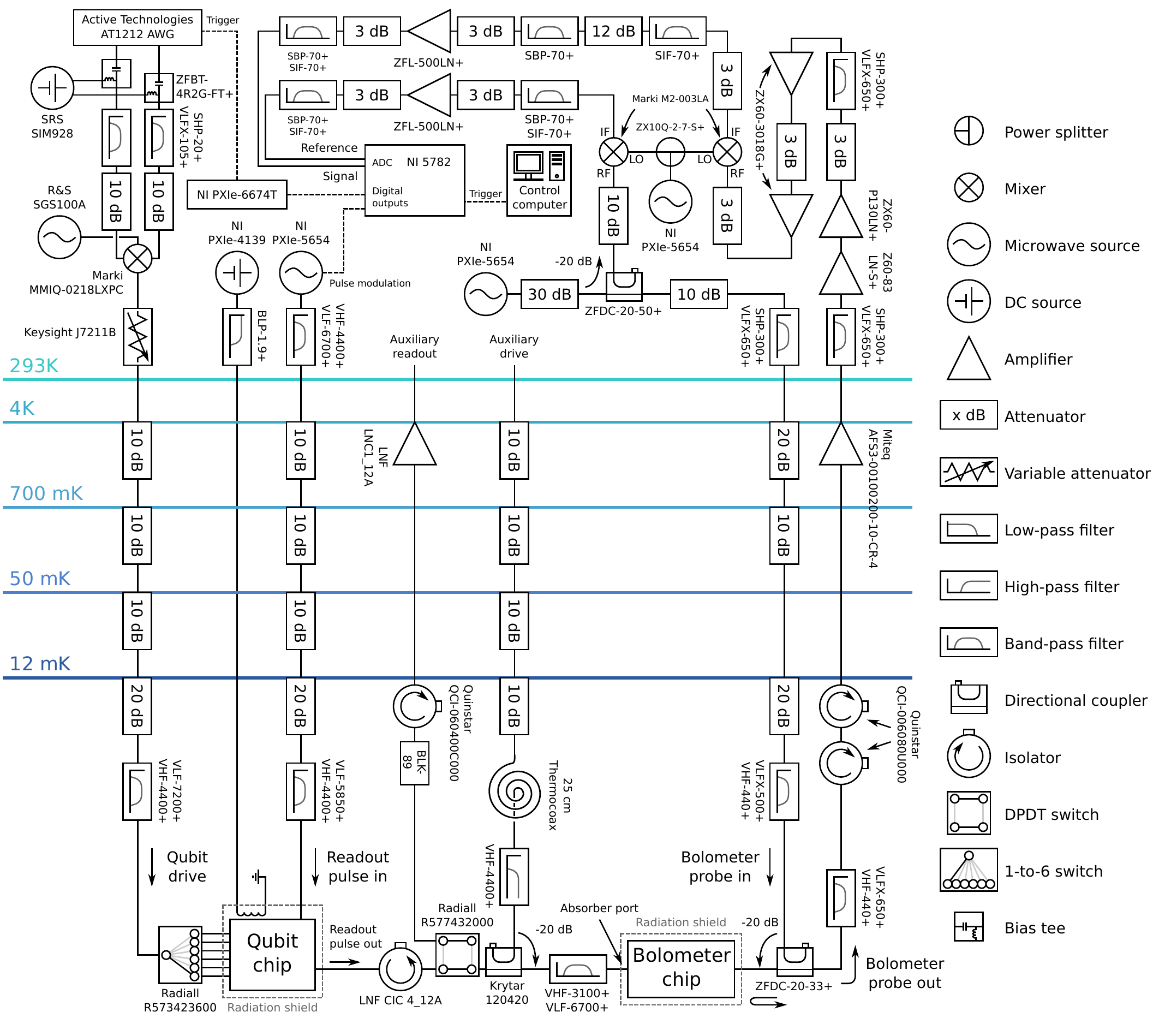}
    \caption{%
        \label{fig:detailed_exp_setup}%
        \textbf{Full experimental setup.}
        The labels for the filters, directional couplers, amplifiers, and the power splitter refer to Mini-Circuits model numbers.
        The band-pass filters with two labels denote a low-pass filter and high-pass filter in series.
    }
\end{figure*}

A diagram of the full experimental setup is shown in Extended Data Fig.~\ref{fig:detailed_exp_setup}.
A PXI control computer initiates a measurement by sending a software trigger to a National Instruments NI-5782 transceiver module connected to a NI PXIe-7962R analog-to-digital converter (ADC) running custom FPGA code.
The transceiver sends further digital trigger signals to initiate pulses to the readout resonator and to excite the qubit.
The clocks of all devices are synchronized using a \SI{10}{\mega\hertz} Rb reference.

The qubit excitation pulse is generated by an Active Technologies AT1212 digital-to-analog converter, upconverted to gigahertz range, and directed into the cryostat via a series of attenuators and filters.
A microwave switch placed at the mixing-chamber plate is used to select which of the six qubits on the chip is being driven.
The qubit frequency is adjusted by applying a magnetic field to the qubit-resonator chip using a hand-wound coil outside the sample holder.

The readout pulse is generated by a microwave source using pulse modulation.
The pulse is reflected off of the resonator on the qubit-resonator chip, after which it is directed through a circulator, a double pole double throw (DPDT) switch, a directional coupler and some filters, before reaching the bolometer absorber port.
The components between the qubit-resonator and bolometer chips are needed only to individually characterize the qubit-resonator chip (by toggling the DPDT switch) and the bolometer (by applying a pulse through the directional coupler), and could be removed in future experiments.

To probe the bolometer, a continuous microwave tone is split into two paths, a reference and a signal.
The signal is directed into the cryostat, reflected off of the probe port of the bolometer, and amplified and demodulated to an intermediate frequency (IF) using a local oscillator (LO), which is detuned from the probe tone by a fixed IF frequency of $\SI{70.3125}{\mega\hertz}$.
The reference is demodulated by the LO without passing through the cryostat, and is used as an amplitude and phase reference for the signal.
Both IF signals are digitized by the NI-5782 at a sampling rate of $\SI{250}{\mega\sample/\second}$ and digitally demodulated into in-phase ($I$) and quadrature ($Q$) components.
The NI-5782 also handles ensemble-averaging the data where applicable, as well as further downsampling by boxcar averaging a variable number of adjacent points.
From time-domain data in the $IQ$-plane, we calculate the average signal before the pulse, $\tilde V_0$, and the average signal during the pulse $\tilde V$, as determined by $t_0$ and $t_{\RO}$.
The data is then rotated in the $IQ$-plane such that $\tilde V - \tilde V_0$ lies on the $I$ axis.
We define $\bar V$ and $\bar V_0$ discussed in the main text as the $I$-components of $\tilde V$ and $\tilde V_0$, respectively, after applying this rotation.
For the single-shot experiments, we apply a common rotation across all shots, with the rotation angle chosen such that the readout fidelity is maximized.
Note that the phase of the signal in the $IQ$ plane is due to the reflection off of the effective $LC$ circuit of the bolometer, and is thus completely independent of the phase of the photons emitted by the readout resonator.

The data acquisition and storage is managed by the QCoDeS~\cite{nielsen2022} data acquisition framework.
Data analysis and fitting is carried out using the NumPy~\cite{harris2020}, SciPy~\cite{virtanen2020}, xarray~\cite{hoyer2017} and lmfit~\cite{newville2022} libraries.

\subsection{Sample characterization}

\begin{figure*}
    \includegraphics[width=\linewidth]{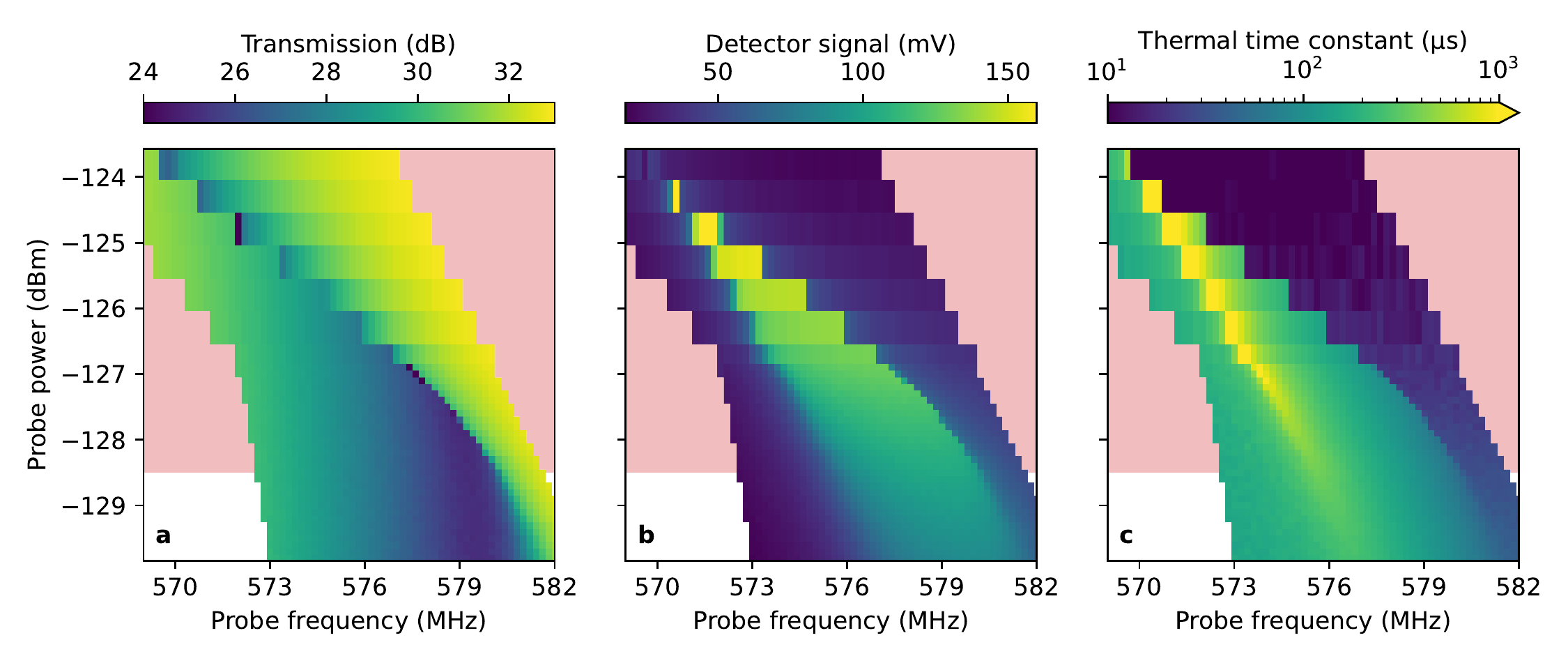}
    \caption{
        \label{fig:bolometer_characterization}%
        \textbf{Bolometer characterization.}
        (a)~Total transmission power divided by the reference power, (b)~detector signal $S$, and (c)~extracted thermal time constant $\tau_{\rb}$ of the bolometer as functions of the probe frequency $f_{\rp}$ and power $P_{\rp}$.
        The red background approximately indicates the power region where the bolometer exhibits bistability.
    }
\end{figure*}

Characterization measurements of the bolometer are shown in Extended Data Fig.~\ref{fig:bolometer_characterization}.
Extended Data Fig.~\ref{fig:bolometer_characterization}(a) shows the signal coming out of the cryostat normalized by the reference, as discussed above.
The data are averaged for \SI{1}{\milli\second} over 16 repetitions.
For each pixel in Extended Data Fig.~\ref{fig:bolometer_characterization}(b), a \SI{2}{\milli\second} pulse is applied to the bolometer absorber port via the readout resonator with an off-resonant tone of $f_{\rd} = \SI{5.4}{\giga\hertz}$ and $P_{\rd} = \SI{-108}{\dBm}$.
From these data, we extract the time constant $\tau_{\rb}$ shown in Extended Data Fig.~\ref{fig:bolometer_characterization}(c) by fitting the time-domain signal to an exponential model.

\begin{figure*}
    \includegraphics[width=\linewidth]{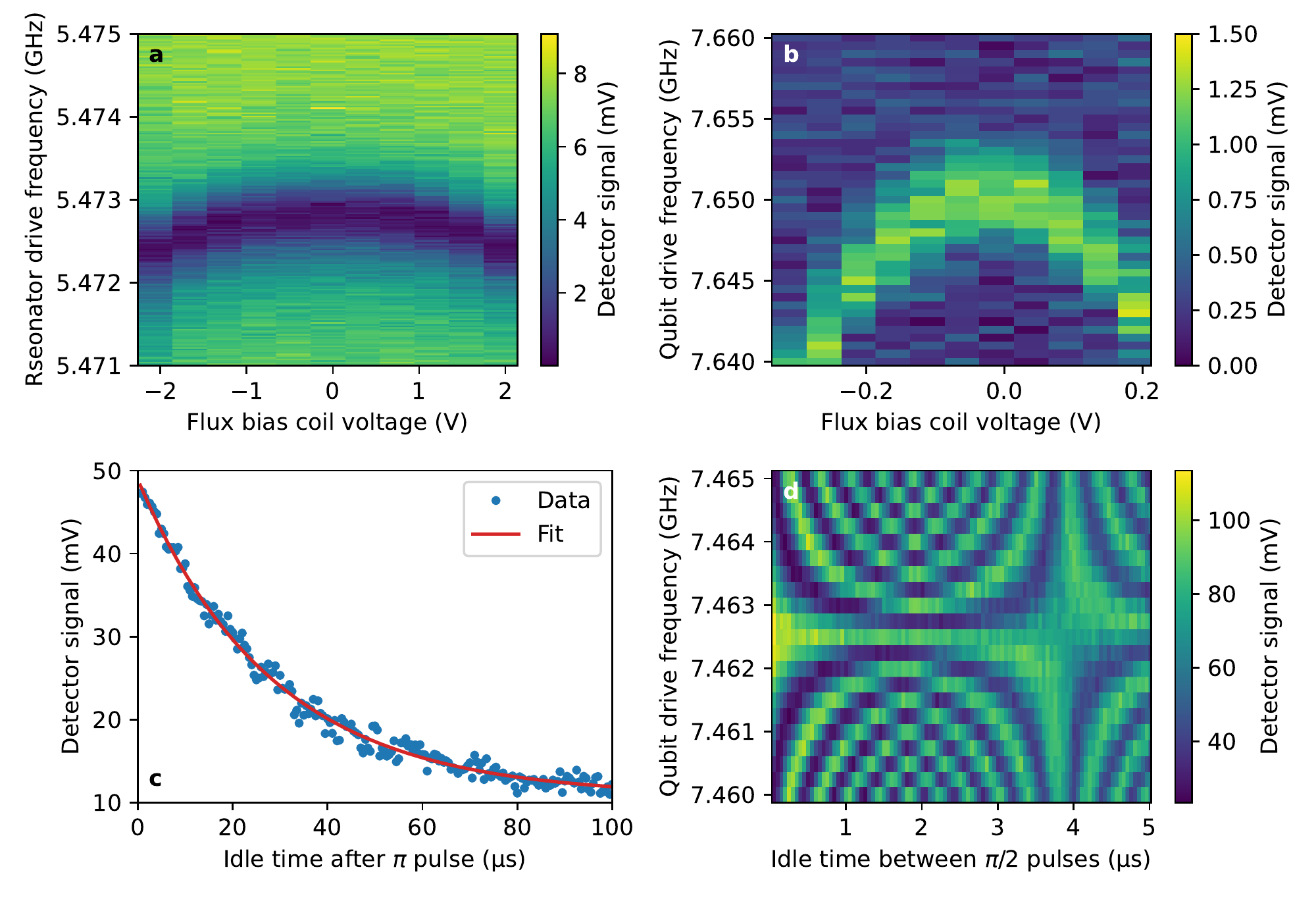}
    \caption{%
        \label{fig:qubit_characterization}%
        \textbf{Qubit characterization using bolometer.}
        (a)~Detector signal $S$ as a function of the bias voltage applied to the flux bias coil and of the frequency of the resonator drive pulse.
        The resonator frequency, indicated by the minimum detector signal at a given flux value, clearly changes as a function of the flux bias.
        The data of panel (a) are averaged over 4 repetitions.
        (b)~Detector signal $S$ as a function of the flux bias voltage and the frequency of a continuous tone applied to the qubit.
        The data is averaged over 32 repetitions. The data of panels (a) and (b) are acquired using a readout power of $P_{\rd} = \SI{-132.8}{\dBm}$ and a pulse duration $t_{\RO} = \SI{1}{\milli\second}$.
        (c)~Detector signal $S$ as a function of the idling time between a $\pi$ pulse applied to the qubit and the readout pulse, providing the $T_1 = \SI{28}{\micro\second}$ of the qubit.
        In these measurements, $P_{\rd} = \SI{-124.8}{\dBm}$ and $t_{\RO} = \SI{20}{\micro\second}$, and the data are averaged over 4096 repetitions.
        (d)~Detector signal as a function of the idle time between the $\pi/2$ pulses and the qubit drive frequency in a Ramsey experiment which is used to extract the $T_2$ of the qubit.
        These data are acquired with $P_{\rd} = \SI{-119.8}{\dBm}$ and $t_{\RO} = \SI{25}{\micro\second}$, averaged over 1024 repetitions.
    }
\end{figure*}

Extended Data Fig.~\ref{fig:qubit_characterization} shows standard characterization experiments of the qubit-resonator system using the bolometer.
Panels (a)--(b) show single-tone and two-tone spectroscopy, respectively, from which we extract an initial estimate for the resonator and qubit frequencies.
In Fig.~\ref{fig:qubit_characterization}(c), a $\pi$ pulse is applied to the qubit, and the readout pulse is applied after different delays to extract the $T_1$ decay time.
Extended Data Fig.~\ref{fig:qubit_characterization}(d) shows the result of a Ramsey experiment, where the qubit is driven by two $\pi/2$ pulses with a varying idle time between them, for different modulation frequencies of the pulse.
These data are used to extract the $T_2$ time, given by the exponentially decaying envelope of the oscillation.
The signal level varies in Extended Data Fig.~\ref{fig:qubit_characterization}, since different values of $P_{\rd}$ and $t_{\RO}$ were used during the characterization while optimal values for these parameters were not yet known.
In addition, the cryostat was thermally cycled between the acquisition of the data of panels (a), (b) and those of (c), (d).
This is why the qubit frequency has decreased in panel (d) compared with panel (b) and Fig.~\ref{fig:example_measurements}(e).

\subsection{Detector signal probability distribution with long averaging time}

Following Ref.~\cite{gambetta2007}, we develop a model for the probability distribution of the detector signal for the qubit nominally prepared in $\ket{\re}$ (with some preparation error) and for the averaging time being significant compared to the energy decay time $T_1$.
Let $v_{\rg/\re}(t, t_{\rd})$ be the time domain signal of the bolometer output when the qubit is in the state $\ket{\rg}$ or $\ket{\re}$ at the beginning of the readout pulse, $t = 0$.
Here, $t_{\rd}$ denotes the time at which the qubit instantaneously decays to $\ket{\rg}$ if it was initially in $\ket{\re}$ ($v_{\rg}$ is independent of $t_{\rd}$).
We assume that no thermal excitation events occur, and that the noise in the signal is Gaussian.
Formally, the signal is given by
\begin{align}
    v_{\rg/\re}(t, t_\rd) \rd t = u_{\rg/\re}(t, t_\rd) \rd t + \sqrt{P_{\rN}}\, \rd W(t)
    ,
\end{align}
where $u_{\rg/\re}(t, t_{\rd})$ is the evolution of the output in the absence of noise, $P_{\rN}$ is the noise power spectral density and $\rd W(t)$ is the Wiener increment~\cite{jacobs2010}. 

To incorporate the bolometer dynamics, we model $u_{\rg/\re}(t, t_{\rd})$ as exponentially approaching some steady-state values $c_{\rg}$ and $c_{\re}$, for the qubit in the state $\ket{\rg}$ or $\ket{\re}$, respectively~\cite{kokkoniemi2019}.
Namely, we assume that the time evolution is of the form
\begin{align}
    u_{\rg}(t) &= c_{\rg} (1 - e^{-t/\tau_{\rb}})
\end{align}
and
\begin{align}
    \label{eq:bolometer_dynamics_e}
    \begin{aligned}
        u_{\re}(t, t_{\rd})
        &=
        \hphantom{{}+{}}
        \theta(t_{\rd} - t)
        c_{\re} \left(1 - \re^{-t/\tau_{\rb}}\right)
        \\
        &
        \hphantom{{}={}}
        +
        \theta(t - t_{\rd})
        \left[
            (c_{\rg} - c_{\re}') \left(1 - \re^{-(t-t_{\rd})/\tau_{\rb}}\right) + c_{\re}'
        \right]
        ,
        \rule[-1.5em]{0pt}{0pt} 
    \end{aligned}
\end{align}
where $\theta(x)$ is the step function and $c_{\re}' = c_{\re}(1 - \re^{-t_{\rd} / \tau_{\rb}})$.

Averaging the bolometer output from $t_0$ to $t_{\RO}$ yields the averaged detector signal
\begin{align}
    \bar V_{\rg/\re}(t_{\rd})
    &= \frac{1}{t_{\RO} - t_0} \int_{t_0}^{t_{\RO}} v_{\rg/\re}(t, t_{\rd}) \rd t
    \nonumber
    \\
    &= \bar U_{\rg/\re}(t_{\rd}) + X[0, \sigma^2]
    ,
\end{align}
where $\bar U_{\rg/\re}(t_{\rd}) \equiv (t_{\RO} - t_0)^{-1}\int_{t_0}^{t_{\RO}} u_{\rg/\re}(t, t_{\rd}) \, \rd t$ and $X[0, \sigma^2]$ is a normally distributed random variable with mean $0$ and standard deviation $\sigma = \sqrt{P_{\rN} / (t_{\RO} - t_0)}$.
Above, we have neglected the constant $\bar V_0$, which simply shifts $\bar V_{\rg/\re}$.
With a fixed $t_{\rd}$, the probability of obtaining a given detector voltage $V$ thus obeys the probability distribution
\begin{align}
    P_{\rg/\re}\left(V|t_{\rd}\right)
    =
    \frac{1}{\sqrt{2\pi \sigma^2}} \exp\left\{-\frac{\left[V-\bar U_{\rg/\re}(t_\rd)\right]^2}{2\sigma^2}\right\}
    ,
\end{align}
which is a Gaussian centered around $\bar U_{\rg/\re}(t_{\rd})$ having a standard deviation $\sigma$ as defined above.

If the qubit is in the state $\ket{\rg}$ at $t = 0$, the probability distribution $P_{\rg}(V)$ is simply a Gaussian centered around the constant value $\bar U_{\rg}$.
When the qubit is initially in $\ket{\re}$, the probability distribution is obtained by calculating the average of $P_{\re}(V|t_{\rd})$ for all possible realizations of the qubit decay time $t_{\rd}$ weighted by the probability density $P(t_{\rd})$ of the decay occurring at $t_{\rd}$:
\begin{align}
    \label{eq:probability_distribution_e}
    P_{\re}(V) &= \int_{0}^{\infty} P(V|t_{\rd}) P(t_{\rd}) \, \rd t_{\rd},
\end{align}
where $P(t_{\rd})$ is exponentially distributed: $P(t_{\rd}) = \re^{-t_{\rd}/T_1}/T_1$.
Thus the total probability distribution for the qubit nominally prepared in $\ket{\re}$ is the following weighted sum of these distributions:
\begin{align}
    \label{eq:probability_distribution_tot}
    P_{\re}^{\text{tot}}(V) = P_{\rx} P_{\rg}(V) + (1 - P_{\rx})P_{\re}(V)
    ,
\end{align}
where $P_{\rx}$ is the probability that the qubit was actually in the state $\ket{\rg}$ at $t = 0$. Here, $P_\rx$ includes state preparation errors, as well as the $T_1$ decay of the qubit during the short delay between the application of the $\pi$ pulse and the start of the readout pulse.

In the case where $u_{\rg/\re}(t, t_{\rd})$ is step-function-like, Eqs.~\eqref{eq:probability_distribution_e} and \eqref{eq:probability_distribution_tot} have analytical expressions~\cite{gambetta2007}.
However, we do not obtain an analytical expression for $P_{\re}(V)$ given the temporal evolution arising from Eq.~\eqref{eq:bolometer_dynamics_e}, and hence, we calculate the integral of Eq.~\eqref{eq:probability_distribution_e} numerically.
Scaling this by the bin width used in Fig.~\ref{fig:singleshot_histogram}(b) produces the prediction for the distribution, with $c_{\rg}$, $c_{\re}$, $T_1$, $\sigma$, and $P_{\rx}$ as the fitting parameters.
For the data of Fig.~\ref{fig:singleshot_histogram}(b), we find $c_{\rg} = \SI{24.7}{\milli\volt}$, $c_{\re} = \SI{182}{\milli\volt}$, $T_{1} = \SI{25.8}{\micro\second}$, $\sigma = \SI{17.4}{\milli\volt}$ and $P_{\rx} = 0.20$.

The readout fidelity is given by
\begin{align}
    \label{eq:fidelity}
    F
    &= 1 - \int_{-\infty}^{V_{\text{th}}} P_{\re}^{\text{tot}}(V) \, \rd V - \int_{V_{\text{th}}}^{\infty} P_{\rg}(V) \, \rd V
    ,
\end{align}
where $V_{\text{th}}$ is the threshold value for assigning the measurement outcome to $\ket{\rg}$ or $\ket{\re}$, and we again assume that no thermal excitations occur.
If we further assume that $P_{\rx} = 0$ and that the qubit never relaxes, $P_{\re}^{\text{tot}}(V)$ reduces to a Gaussian distribution.
In this case, the readout fidelity is maximized with $V_{\text{th}} = (\bar U_{\re} + \bar U_{\rg})/2$, which yields
\begin{align}
    \label{eq:fidelity_snr}
    F &= \operatorname{erf}\left(\frac{\bar U_{\re} - \bar U_{\rg}}{2\sqrt{2\sigma^2}}\right)
    ,
\end{align}
where $\operatorname{erf}(x)$ is the error function.

\subsection*{Estimate of parameters needed for 99.9\% fidelity}

Here, we quantitatively discuss how to achieve 99.9\% fidelity with bolometric readout following the improvements to the experiment proposed in the main text.
To be able make a fair comparison between the performance of different bolometers presented in the literature~\cite{kokkoniemi2019, kokkoniemi2020}, we carry out this analysis by assuming that we operate in the calorimetric mode. Namely, we interpret the distributions of Fig.~\ref{fig:singleshot_histogram}(b) to represent a measurement of the energy of the absorbed pulses. In such an experiment, the signal is directly proportional to the energy packet arriving at the bolometer whereas the width of the distribution is independent of the energy.

Let us first assume that the length of the readout pulse is reduced to 200~ns without changing the readout power. This means that the bolometer-absorbed energy, and hence the SNR in Fig.~\ref{fig:singleshot_histogram}(b) is reduced by a factor of 70. By changing the metallic bolometer to the graphene bolometer, the energy resolution has been observed to increase by a factor of 13~\cite{kokkoniemi2020}. Thus to obtain 99.9\% fidelity, one needs to increase the amount of energy absorbed by the graphene bolometer in $\SI{200}{\nano\second}$ by a factor of $ A_{99.9\%}\approx 1.8\times70/13\approx 10$, where the factor of 1.8 arises from the fact that in order to reduce the overlap infidelity of Fig.~\ref{fig:singleshot_histogram}(b) from 7\% to 0.1\%, one needs a factor of 1.8 improvement in the SNR in Eq.~\eqref{eq:fidelity_snr}.

Below, we aim to show that by making different improvements to the measurement scheme, we can arrive at several factors $A_\alpha$ of increment to the energy absorbed by the bolometer, and taken all these factors into account, it is possible to exceed the required SNR, i.e., $A_{99.9\%}=10<\prod_\alpha A_\alpha$. Since we have assumed $\SI{200}{\nano\second}$ readout time, having a qubit $T_1$ of the order of $\SI{100}{\micro\second}$ is sufficient to reach 99.9\% readout fidelity.
Note that the bolometer is a power sensor, and thus the SNR is directly proportional to the absorbed power for low enough powers considered here~\cite{kokkoniemi2020} [see also Fig.~\ref{fig:example_measurements}(d)].


Firstly, by switching to a transmission-type setup, where all of the readout photons are directed into the bolometer instead of half of them escaping through the input, increases the power incident on the bolometer by a factor of $A_{\text{t}} = 2$.
Secondly, removing additional components between the qubit and bolometer chips reduces losses approximately by $A_{\text{c}} \approx \SI{1}{\dB} \approx 1.25$.
Thirdly, by optimizing the dispersive shift $\chi$ for photodetection-type readout, we expect that the photon number occupying the resonator during readout and thus incident power can be increased by a factor $A_\chi = 2$ based on Ref.~\cite{nesterov2020}.
Fourthly, switching the qubit-resonator system to a configuration where the resonator resonance frequency lies above the qubit frequency (keeping the detuning $|f_{\rr,\rg} - f_{\qq}|$ constant) allows driving with a larger number of photons~\cite{cohen2022}, which we estimate to be at least $A_{\text{a}} = 1.5$ times greater than with our current parameters.
Fifthly, increasing the admissible drive power by doubling the resonator drive frequency to $\SI{10}{\giga\hertz}$ may further double the photon energy $h f_{\rr, \rg}$, introducing a factor $A_{2f} = 2$.
In total, the improvement to the SNR is given by
\begin{equation}
    A_{\text{t}} A_{\text{c}} A_{\text{a}} A_{2f} A_\chi
    \approx
    15
    > A_{99.9\%}
    ,
\end{equation}
and thus a high fidelity is feasibly achievable.

\section*{Data availability}

The data that support the findings of this study are available at \url{https://doi.org/10.5281/zenodo.7773981}.

\section*{Author contributions}

A.M.G. and S.N. conducted the experiments and analysed the data.
W.L. designed and fabricated the bolometer chip.
The qubit chip was designed by S.K. and fabricated by J.M. with help from V.~Ve.
G.C., P.S. and Q.C. assisted with characterizing and operating the bolometer.
A.M.G. and V.~Va. developed the model for the signal probability distribution.
The manuscript was written by A.M.G. and M.M., with comments from all authors.
The work was conceived and supervised by M.M.

\section*{Acknowledgements}
The authors acknowledge funding from the Academy of Finland Centre of Excellence program (project nos. 352925, 336810, 336817, and 336819), European Research Council under Advanced Grant no. 101053801
(ConceptQ), the Future Makers Program of the Jane and Aatos Erkko Foundation, the Technology Industries of Finland Centennial Foundation and the Finnish Foundation for Technology Promotion.
We thank Arman Alizadeh, Slawomir Simbierowicz, and Russell Lake for useful discussions.
%
%

\section*{Competing interests}
M.M. declares that that he is a Co-Founder of the quantum-computer company IQM Finland Oy. Other authors declare no competing interests.

\bibliography{qubit-readout-using-bolometer-bibliography} 

\end{document}